\documentclass[aps,prb,groupedaddress,twocolumn,10pt,showpacs]{revtex4-1}
\usepackage[pdftex]{graphicx}
\usepackage{amsmath,amssymb,color,subfigure}

\newcommand{\be}{\begin{equation}}
\newcommand{\ee}{\end{equation}}
\newcommand{\bea}{\begin{eqnarray}}
\newcommand{\eea}{\end{eqnarray}}
\renewcommand{\vec}[1]{\mathbf{#1}}

\newcommand{\lab}{\langle}
\newcommand{\rab}{\rangle}

\newcommand{\kx}{\langle - k_x \rangle}

\renewcommand{\Re}{\mathop{\mathrm{Re}}}
\renewcommand{\Im}{\mathop{\mathrm{Im}}}

\begin{document}
\title{Dynamical Conductivity Across The Disorder-Tuned Superconductor-Insulator Transition}
\author{Mason Swanson$^{1}$}
\author{Yen Lee Loh$^{2}$}
\author{Mohit Randeria$^{1}$}
\author{Nandini Trivedi$^{1}$}
\affiliation{(1) Department of Physics, The Ohio State University, Columbus, OH  43210, USA}
\affiliation{(2) Department of Physics and Astrophysics, University of North Dakota, Grand Forks, ND  58202, USA}
\date{
\today}

\begin{abstract}
We calculate the dynamical conductivity $\sigma(\omega)$ and the bosonic (pair) spectral function $P(\omega)$ 
from quantum Monte Carlo simulations across clean and disorder-driven superconductor-insulator transitions (SIT). 
We identify characteristic energy scales in 
the superconducting and insulating 
phases that vanish at the transition due to enhanced quantum fluctuations, despite the persistence of a robust fermionic gap across the SIT. Disorder leads to enhanced absorption in $\sigma(\omega)$ at low frequencies compared to the SIT in a clean system. 
Disorder also expands the quantum critical region, due to a change in the universality class, with an underlying $T=0$ critical point with a universal low-frequency conductivity $\sigma^\ast \simeq 0.5 (4e^2/h)$.
\end{abstract}
\maketitle


The interplay of superconductivity and localization has proven to be a rich and intriguing problem, especially in two dimensions~\cite{hebard1990,shahar1992,adams2004,Sambandamurthy2004,steiner2005,stewart2007,gantmakher2010}.
Both paradigms stand on the shoulders of giants -- the BCS theory of superconductivity and the Anderson theory of localization.  Yet, when the combined effects of superconductivity and disorder are considered, both paradigms break down, even for s-wave superconductors. 

It has been shown~\cite{ghosal1998,ghosal2001,bouadim2011} 
in model fermionic Hamiltonians with attraction between electrons and disorder arising from random potentials,
that the single-particle density of states continues to show a hard gap 
across the disorder-driven quantum phase transition and that pairs continue to survive into the insulating state. 
The superconducting transition temperature $T_c$, however, does decrease with increasing disorder 
and vanishes at a critical disorder signaling a superconductor-insulator transition (SIT). 
These theoretical predictions are supported by scanning tunneling spectroscopy experiments~\cite{sacepe2010,sacepe2011,mondal2011,sherman2012} 
and by magnetoresistance oscillations~\cite{stewart2007} in disordered thin films.

Recent conductivity measurements at frequencies well within the superconducting gap 
($0$--$20$ GHz)~\cite{crane_fluctuations_2007,crane_survival_2007,weiliu2011,weiliu2012,Lemarie2013, mondal2013} 
have observed low-frequency features that cannot be accounted for by pair-breaking mechanisms. 
A theoretical understanding of the low-frequency dynamical conductivity is vital for understanding the role of fluctuations and for 
guiding future experiments that probe the SIT.

\begin{figure}[!t]
\centerline{\includegraphics[width= 0.48 \textwidth]{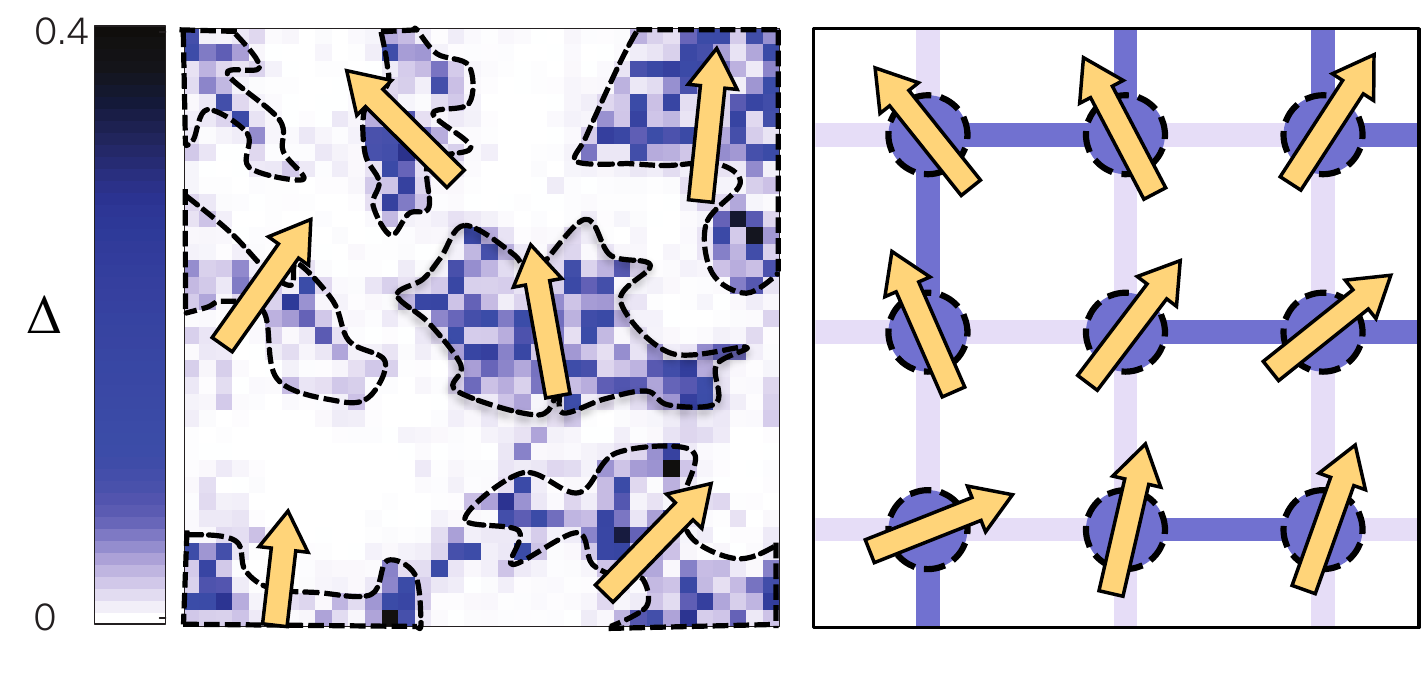}}
\caption{	
The emergent inhomogeneity of  the local pairing amplitude $\Delta(\vec{r})$ in a disordered superconductor in the left panel 
and the robustness of the single particle gap~\cite{ghosal1998,ghosal2001,bouadim2011}
across the SIT suggests an effective low-energy description in terms of a disordered quantum XY model shown on the right. 
The quantum phase transition occurs when long range phase coherence is lost between weakly connected ``superconducting islands'' tuned by the ratio $E_c/E_J$ of charging energy to Josephson coupling as well as by disorder, modeled by removing a fraction $p$ of the Josephson bonds.
\label{fig:model}
	}
\end{figure}

The robustness of the single-particle gap across the SIT suggests that the low-energy physics near the SIT can be described by 
an effective ``bosonic'' Hamiltonian, the disordered quantum XY model, where the relevant degrees of freedom are
the phases of the local superconducting order parameter.
This model is also relevant for ultracold atomic gases in optical lattices where the transition is tuned by changing the tunneling of bosons 
compared to their on-site repulsion~\cite{fisher1989, jaksch1998, greiner2002, spielman2007}.  More recently, it has also become possible to include disorder in optical lattices using speckle patterns.  By increasing the strength of the disorder potential it could be possible to drive quantum phase transitions from a superfluid to a Bose glass~\cite{gurarie2009,billy2008,kondov2011,kondov2013}; our results are also relevant for such experiments.

We map the quantum (2+1)D XY Hamiltonian to an anisotropic classical 3D XY model~\cite{cha1991,sorensen1992,cha1994} and
simulate the model using Monte Carlo methods.
We focus on the behavior of two dynamical quantities of fundamental significance, the conductivity
$\sigma(\omega)$ and the boson (``pair'') spectral function $P(\omega)$ 
obtained by analytic continuation from 
imaginary time using the maximum entropy method supplemented by sum rules.  
Disorder is introduced into the quantum model by breaking bonds (``Josephson couplings'') on a 2D square lattice with a probability $p$.
We compare the results of the disorder-driven SIT with the clean system~\cite{cha1991, smakov2005}, where the SIT is
tuned by $E_c/E_J$, the charging energy relative to the Josephson coupling.

\begin{figure}[!t]
\centerline{\includegraphics[width= 0.33 \textwidth]{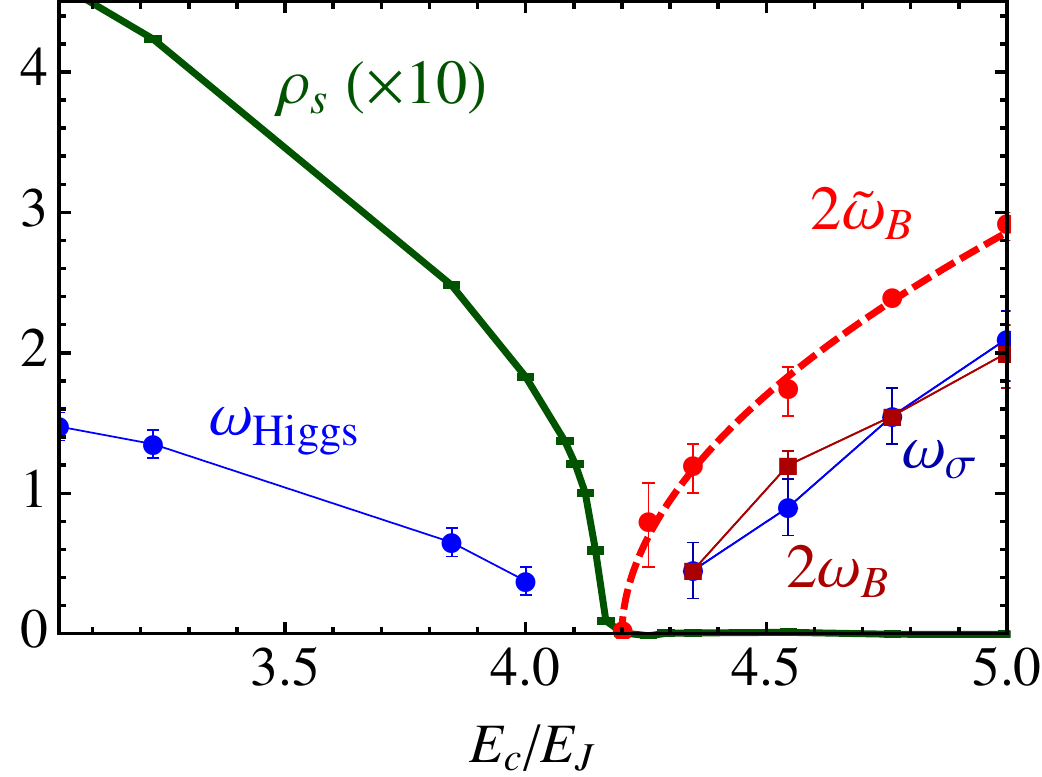}
}
\caption{Energy scales, in units of $E_J$, as a function of the control parameter
$E_c/E_J$ in the clean system. From the SC side, the
superfluid stiffness $\rho_s$ and the Higgs ``mass'' $\omega_{\rm Higgs}$, and from the insulating side,
the optical conductivity threshold $\omega_\sigma$ and the 
boson energy scales $\omega_B$ and $\widetilde{\omega}_B$, vanish at the transition creating a fan-shaped region where quantum critical fluctuations dominate.
}
\label{fig:energyscales}
\end{figure}

Our main results are as follows.

\noindent (1) The conductivity $\Re\sigma(\omega)$ in the clean superconductor 
shows absorption above a threshold  $\omega_{\rm Higgs}$ that can be associated with the scale of the Higgs (amplitude) mode. 
As we approach the SIT from the superconducting (SC) side, both the superfluid stiffness $\rho_s$  
and the Higgs scale $\omega_{\rm Higgs}$ go soft and vanish at the SIT, even though
the fermionic energy gap remains finite across the transition.

 \noindent (2) In the insulating state of the clean system, we find a threshold $\omega_\sigma$
 for absorption in $\Re\sigma(\omega)$ and show that it is twice the gap $\omega_B$ in the bosonic spectral function
 $\Im \, P(\omega)/\omega$. We show that both these scales go soft on approaching the SIT from the insulating side.  
Furthermore, in the insulator, $\Im\sigma(\omega)$ becomes negative at low frequencies, indicating ``capacitive'' response.

\noindent (3) The low-frequency spectral weight in $\Re\sigma(\omega)$ for the disordered system is greatly enhanced 
relative to its clean counterpart, so that there is no clear optical gap in the vicinity of the SIT, despite the existence
of a non-zero fermonic energy gap. We find that enhanced quantum phase fluctuations and rare regions
generate low-frequency spectral weight for $\omega$ well below the clean $\omega_{\rm Higgs}$ scale in the SC state, 
and well below the clean $\omega_\sigma = 2 \omega_B$ scale on the insulating side.

\noindent (4) The spectral function $\Im \, P(\omega)/\omega$ has a characteristic peak in the insulator,
whose energy $\widetilde{\omega}_B$ is a measure of the inverse coherence time scale for bosonic (pair) excitations. 
The vanishing of the superfluid stiffness $\rho_s$ on the SC side and the vanishing of $\widetilde{\omega}_B$ from the
insulating side are shown to demarcate the quantum critical regime at the SIT for both the clean and the disordered system.

\noindent (5) The low-frequency conductivity  $\sigma^\ast$ in the quantum critical regime between the SC and the insulator can be
estimated meaningfully from the integrated spectral weight over a frequency range of the order of the temperature (see Eq.~\ref{sigma-star}).
We find $\sigma^\ast \simeq 0.5 (4e^2/h)$ at the disorder-driven SIT in comparison to $\sigma^\ast \simeq
0.4 (4e^2/h)$ at the SIT in the pure system, in good agreement with 
recent studies~\cite{smakov2005,gazit2013arxiv,witczak2013,chen2013universalconductivity}
of the disorder-free problem.

\begin{figure}[!t]
\centerline{\includegraphics[width= 0.5 \textwidth]{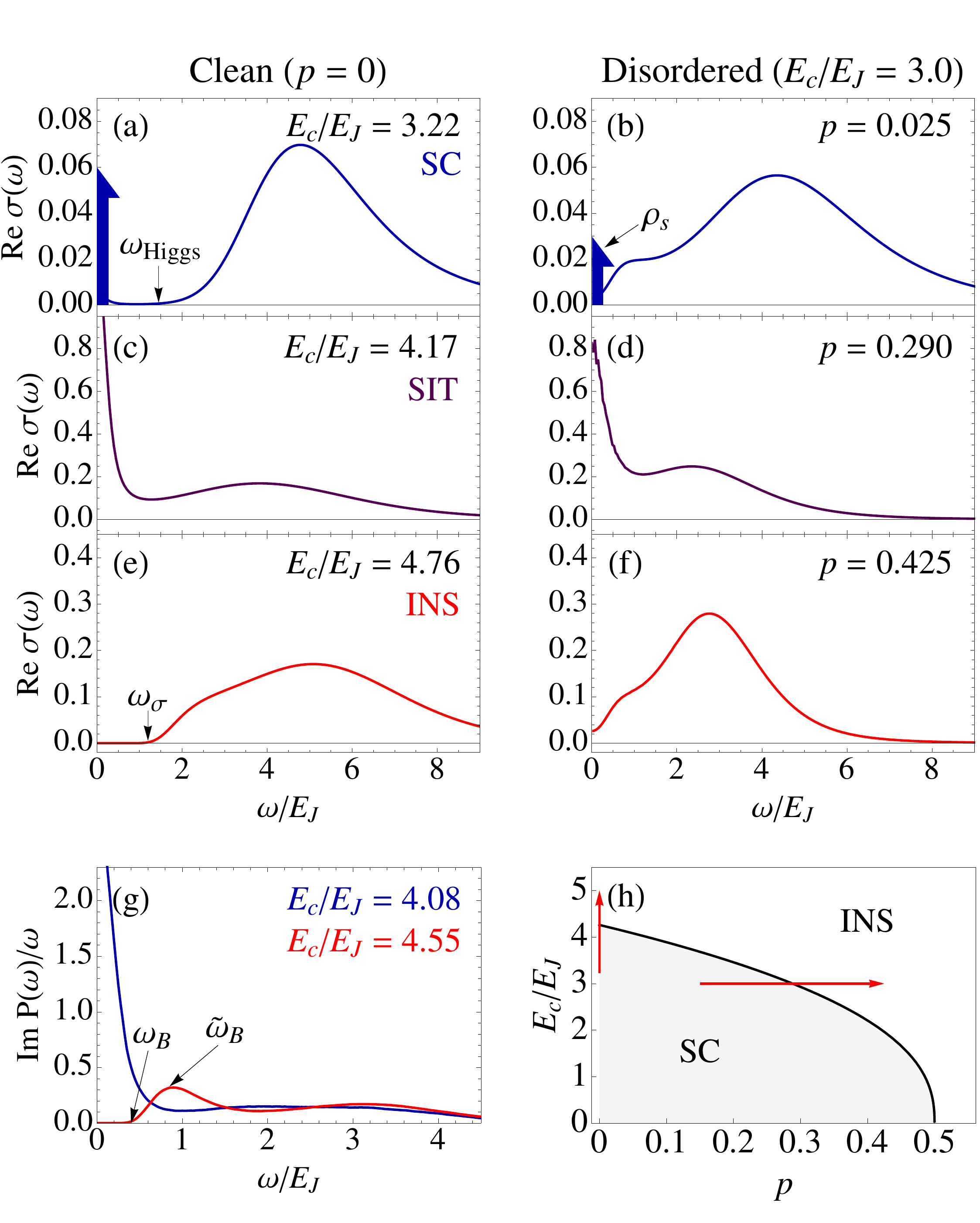}
}
\caption{	
	(a-f): $\Re\sigma(\omega)$ across the (a,c,e) clean ($p\!=\!0$) and (b,d,f) 
	disorder-tuned (fixed $E_c/E_J$) superconductor-insulator transitions.
	(g): Boson spectral function $\Im P(\omega)/\omega$ for a clean superconducting (blue) and insulating (red) state.
	The energy scales shown in Fig.~\ref{fig:energyscales} are indicated in (a-g). 
	All quantities are at fixed temperature $T/E_J = 0.156$, and fixed system size, $256\times256$ for the clean case and $64\times64$ for the disordered.
	In the disordered system, the spectral functions are marked by a significant increase in low frequency weight, 
	obscuring the gap scales of the clean system.
	(h): Schematic phase diagram showing how the SIT can be crossed by either increasing $E_c/E_J$ 
	or by tuning the disorder $p$.}
\label{fig:dynamics}
\end{figure}

\begin{figure*}[!t]
\centerline{\includegraphics[width= 0.275 \textwidth]{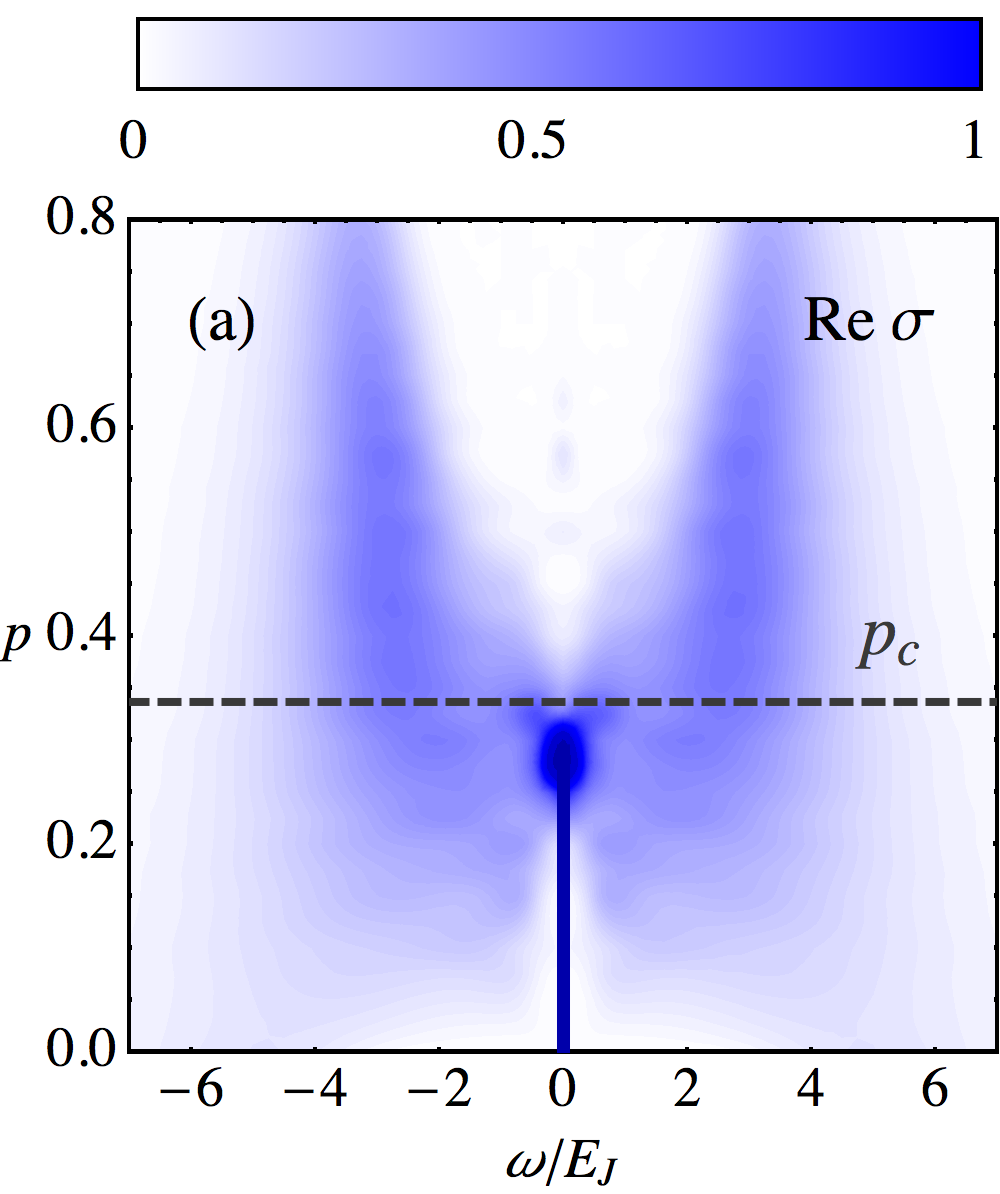} \hspace{0.5 cm}
\includegraphics[width= 0.275 \textwidth]{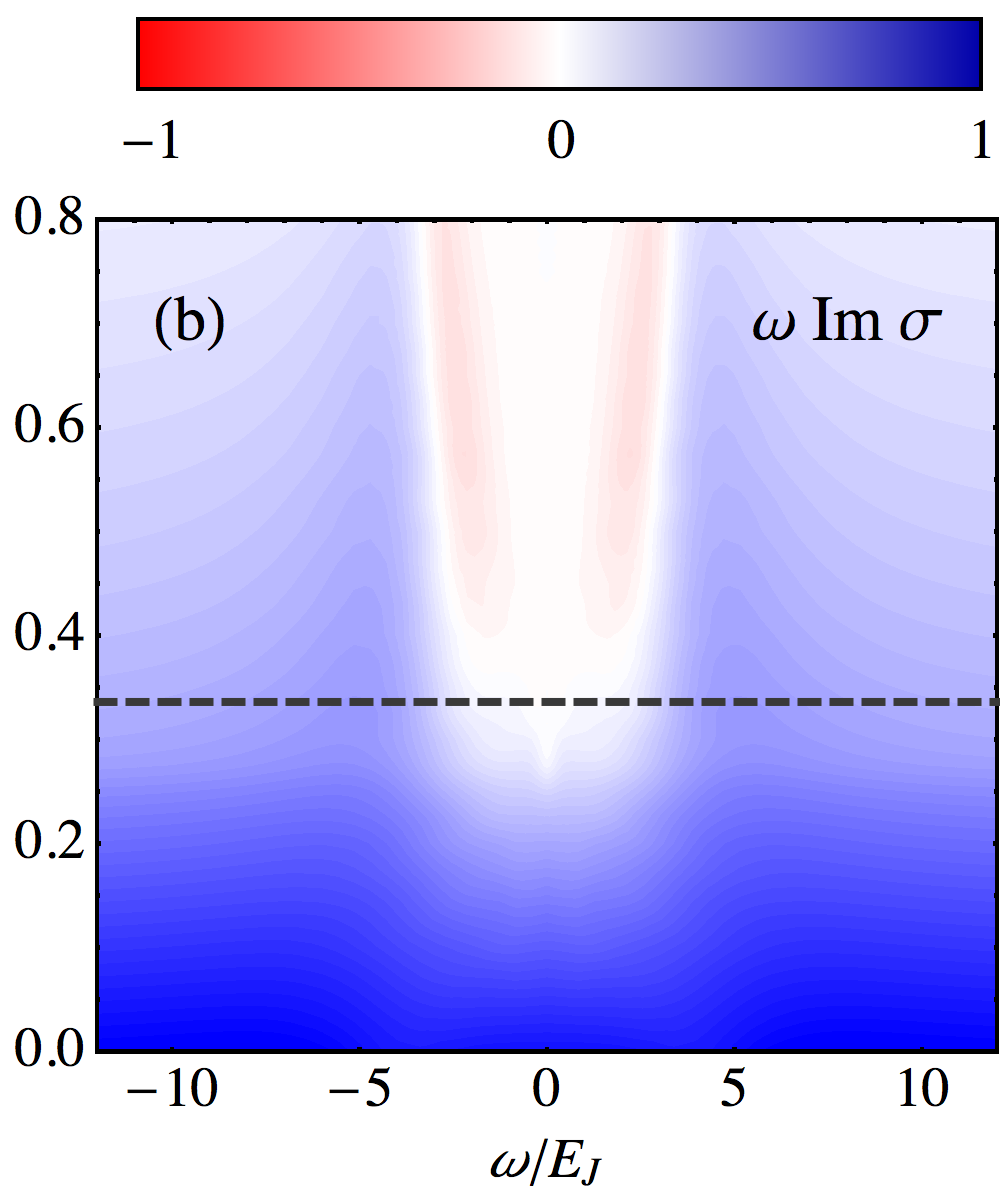} \hspace{0.5 cm}
\includegraphics[width= 0.275 \textwidth]{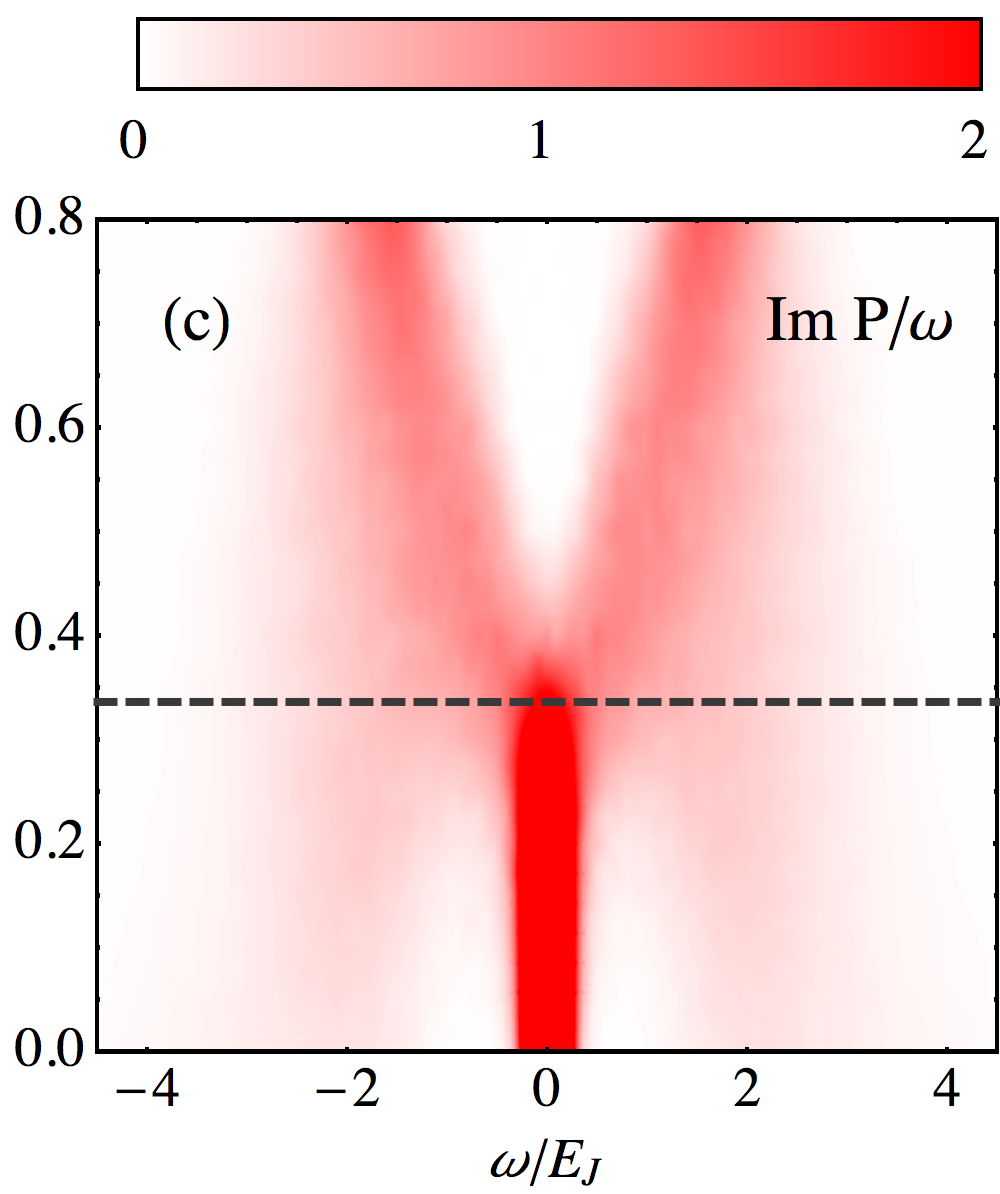}}
\caption{	
	Dynamical response functions across the disorder-tuned SIT.
	The critical disorder $p_c = 0.337$ is marked as a dashed line; $T/E_J = 0.156$, $E_c/E_J = 3.0$ and $L = 64$.
	(a) In the conductivity $\Re \sigma(\omega)$ the superfluid response is evident as a zero-frequency delta function
	of strength $\rho_s$. Deep in the insulator there is a gap in $\Re  \sigma(\omega)$ that grows with disorder.
	(b) $\omega\Im \sigma(\omega)$ shows a crossover from ``inductive'' ($\omega\Im \sigma(\omega) = \rho_s > 0$)
	to ``capacitative'' ($\omega \Im \sigma(\omega) < 0$) behavior at small $\omega$ across the transition.
	(c) The boson spectral function $\Im P(\omega)/\omega$, which has a peak centered about 
	zero frequency in the superconductor, develops a characteristic scale $\widetilde{\omega}_B$ in the insulator that grows with disorder. }
\label{fig:MEMresults}
\end{figure*}

\noindent
{\bf Model}: The quantum XY model is equivalent to a Josephson-junction array, with the Hamiltonian 
\be
\hat{H}_{J} = \frac{E_c}{2} \sum_i{\hat{n}_i}^2 - \sum_{\lab i j \rab} J_{ij} \cos{(\hat{\theta}_i- \hat{\theta}_j)}
\label{ham}
\ee
where the number operator $\hat{n}_i$ at site $i$ is canonically conjugate to the phase operator $\hat{\theta}_i$.
Here $E_c$ is the charging energy.  The Josephson couplings are
$J_{ij}=E_J$ with probability $(1-p)$ and $J_{ij}=0$ with probability $p$. 
The clean system ($p=0$) is a coherent superconductor 
when $E_J$ dominates over $E_c$, with phases aligned across all the junctions. 
However, large $E_c/E_J$ favors a well-defined number eigenstate, 
leads to strong phase fluctuations, and drives the system into an insulating state.
Thus $E_c/E_J$ can be used to tune across the SIT in the clean system.
A quantum phase transition can also be induced by increasing disorder $p$ 
(bond dilution) for fixed $E_c/E_J$. (Fig.~\ref{fig:dynamics}(h)).
Thus Eq.~\ref{ham} is a simple yet non-trivial model that describes a disorder-tuned SIT
with a dynamical exponent $z=1$.


Our results are obtained from calculations of
the superfluid stiffness $\rho_s$, the complex conductivity $\sigma(\omega)$, and the 
boson spectral function $\Im P(\omega)$.
We estimate the superfluid stiffness $\rho_s$ using 
 $\rho_s/\pi= \Lambda_{xx}(q_x\!\rightarrow\!0, q_y\!=\!0, i\omega_n\!=\!0) -\Lambda_{xx}(q_x\!=\!0, q_y\!\rightarrow\!0, i\omega_n\!=\!0)$,
 which is the difference of the longitudinal and transverse pieces of the current-current correlation function $\Lambda_{xx}$.
Here  $j_x(\vec{r},\tau) \sim \sin{[\theta(\vec{r} + \hat{x},\tau) - \theta(\vec{r} ,\tau)]}$ is the current and
$\omega_n = 2\pi n T$ are Matsubara frequencies. 

We use the Kubo formula for the complex conductivity $\sigma(\omega)$ expressed in
terms of $\Lambda_{xx}({\bf q}=0,\tau)$ and transform the imaginary-time QMC results
to real frequency using the maximum entropy method (MEM); see Appendix B.
We have checked our results extensively using sum rules and compared the MEM results
with direct estimates in imaginary time, as described in detail below.
Similarly, we use QMC methods to calculate the imaginary time correlation function 
$P(\mathbf{r},\tau) = \langle a^\dagger(\mathbf{r},\tau)a(0,0) \rangle$, 
where the bosonic creation operator is $a^\dagger = \exp{ i \theta(\mathbf{r},\tau) }$,
and we obtain the spectral function $\Im P(\omega)$ using the MEM.

\noindent
{\bf Superconductor:} 
We first discuss the SC and insulating state in both the clean and disordered systems, before turning to
the quantum critical point.
The SC state is characterized by a non-zero superfluid stiffness $\rho_s$ (see Fig.~\ref{fig:energyscales}).
We use our calculated $\rho_s$ to test the sum rule for the MEM-derived optical conductivity.
The total spectral weight is given by $\int_{0}^{\infty} d\omega \ \Re \sigma(\omega) = {\pi} \langle-k_x\rangle / 2$, where $\langle-k_x\rangle$ is the kinetic energy. 
We find that $\int_{0^+}^{\infty} d\omega \ \Re \sigma(\omega)$
(note the lower limit of $0^+$) calculated from the MEM result differs from $\langle-k_x\rangle$
by an amount that is exactly accounted for by the delta function  $\rho_s \delta(\omega)$.
We have checked this sum rule both in the clean and the disordered systems (see
Appendix B).

In the clean superconductor (Fig.~\ref{fig:dynamics}(a)), $\Re \sigma(\omega)$
shows finite spectral weight above a threshold. 
Note that in the bosonic model, 
the cost of making electron-hole excitations
 is essentially infinite (i.e., much larger than all scales of interest).
Phase fluctuations of the order parameter, $\Psi = A\exp(i\theta)$, lead to a current ${\bf j} \sim \Im \Psi^*\nabla\Psi \sim |A|^2\nabla\theta$.
This then leads to the absorption threshold~\cite{lindner2010,gazit2013} for creating a massive amplitude excitation (Higgs mode) and a massless
phase excitation (phonon). Hence, we identify the threshold in $\Re \sigma(\omega)$ with the Higgs scale $\omega_\text{Higgs}$.
We emphasize that even though the microscopic model (\ref{ham}) has only phase degrees of freedom, 
its long-wavelength behavior upon coarse-graining contains both amplitude (Higgs) and phase fluctuations (phonons and vortices).
In addition, one can show that  $\Re \sigma(\omega)$ has a $\omega^5$ tail at low energies arising from three-phonon absorption in
a clean SC. The large power-law suppression, together with a very small numerical prefactor~\cite{podolsky2011}, 
however, makes this spectral weight too small to be visible in our numerical results for $\Re \sigma(\omega)$. 

\begin{figure}[!htb]
	\centering
	\includegraphics[width= 0.45 \textwidth]{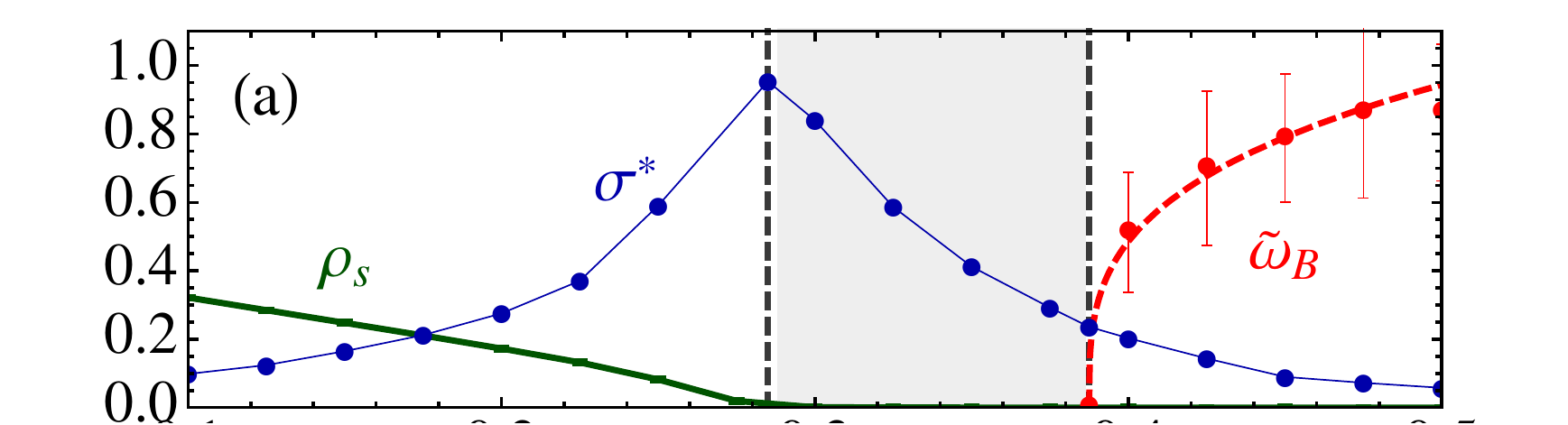}
	\includegraphics[width= 0.45 \textwidth]{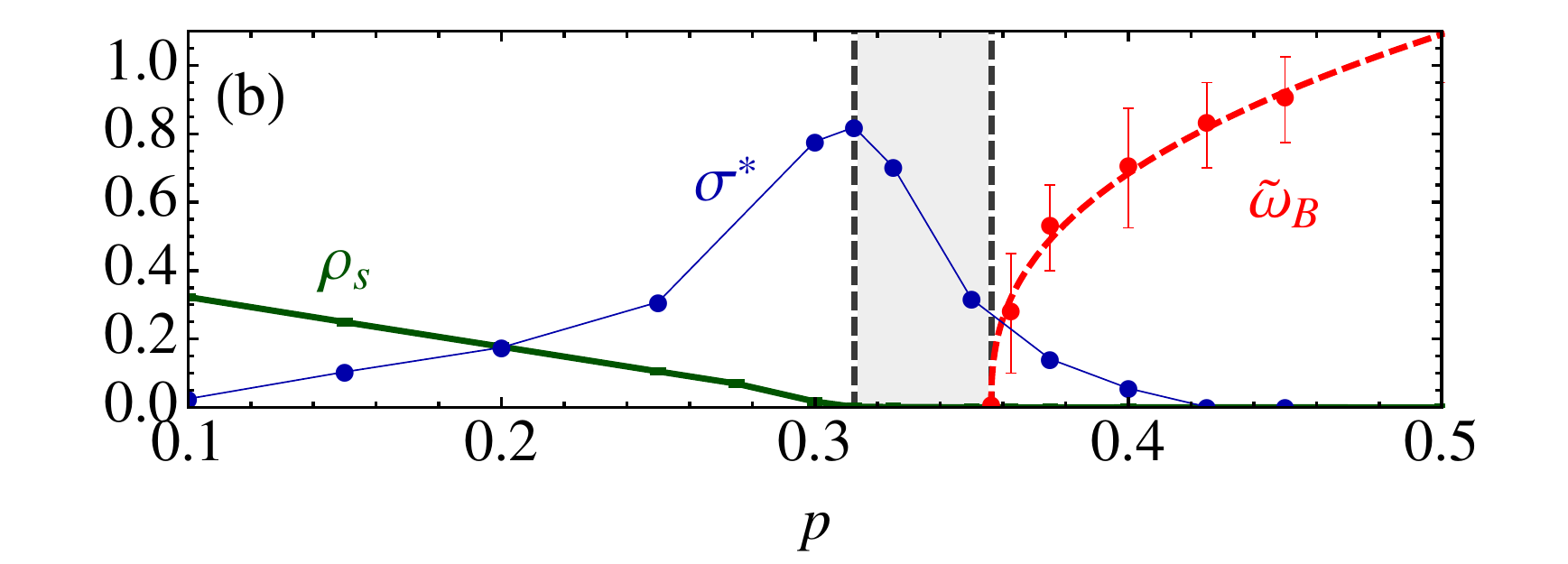}
	\includegraphics[width= 0.45 \textwidth]{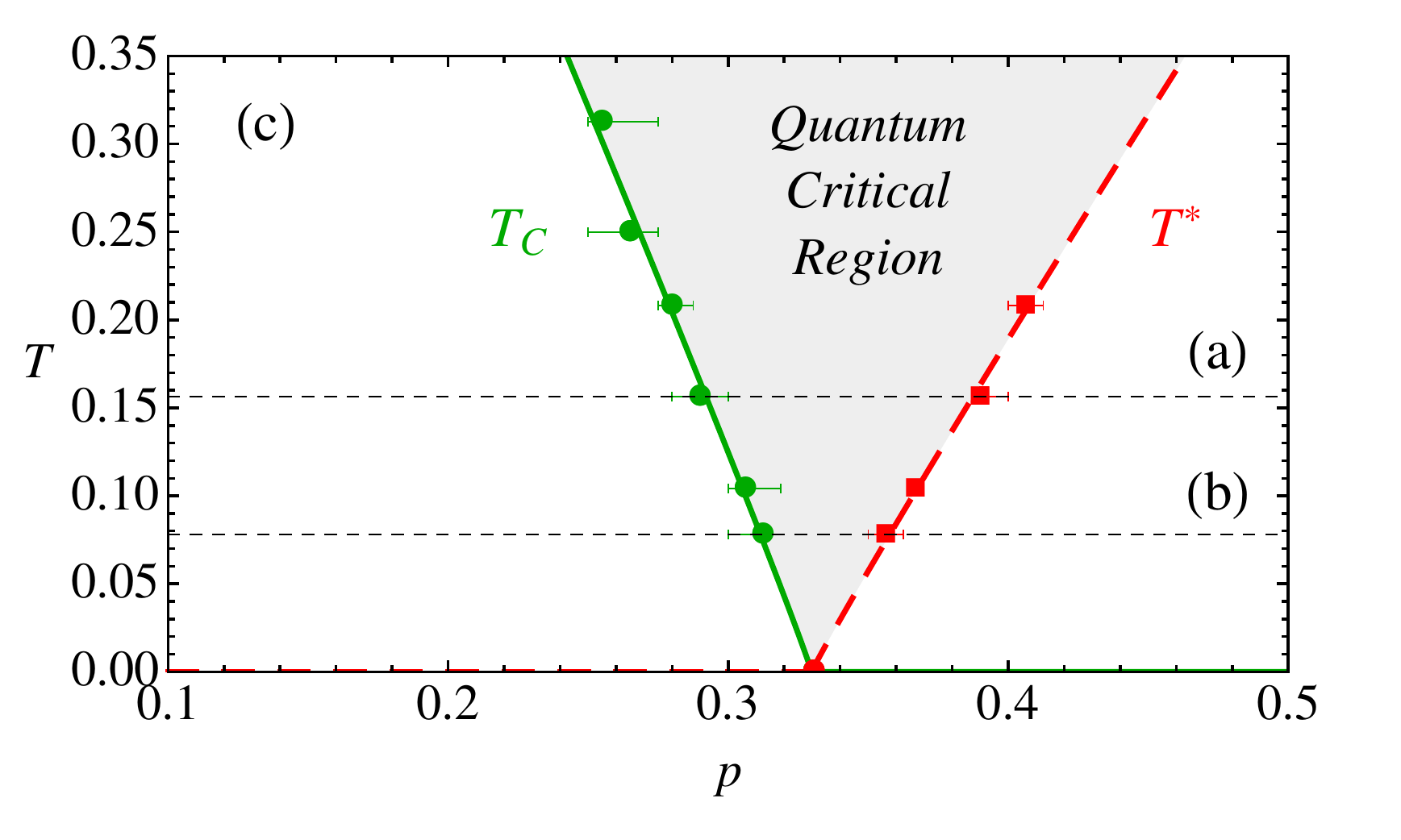}
	\caption{(a,b) Superfluid stiffness $\rho_s$ (green), bosonic scale $\widetilde{\omega}_{\rm B}$ (red) in the insulator, and 
	low-frequency conductivity $\sigma^\ast$ (blue), defined in the text, as functions of disorder $p$ at two different temperatures shown in panel (c).
	The quantities are in unites of $E_J$ and $\sigma_Q = 4e^2/h$, respectively.
	The quantum critical region is shaded gray in all three panels. 
      (c) Phase diagram with $T_c$ determined by vanishing of $\rho_s$ and $T^*$ by the vanishing of $\widetilde{\omega}_{\rm B}$.
      The lines are fits to $|p-p_c|^{z\nu}$ with $p_c \approx 0.337$ and $z\nu \approx 0.96$.}
		\label{QCP}
		\end{figure}
		\begin{figure}[!htb]
\centerline{
\includegraphics[width= 0.22 \textwidth]{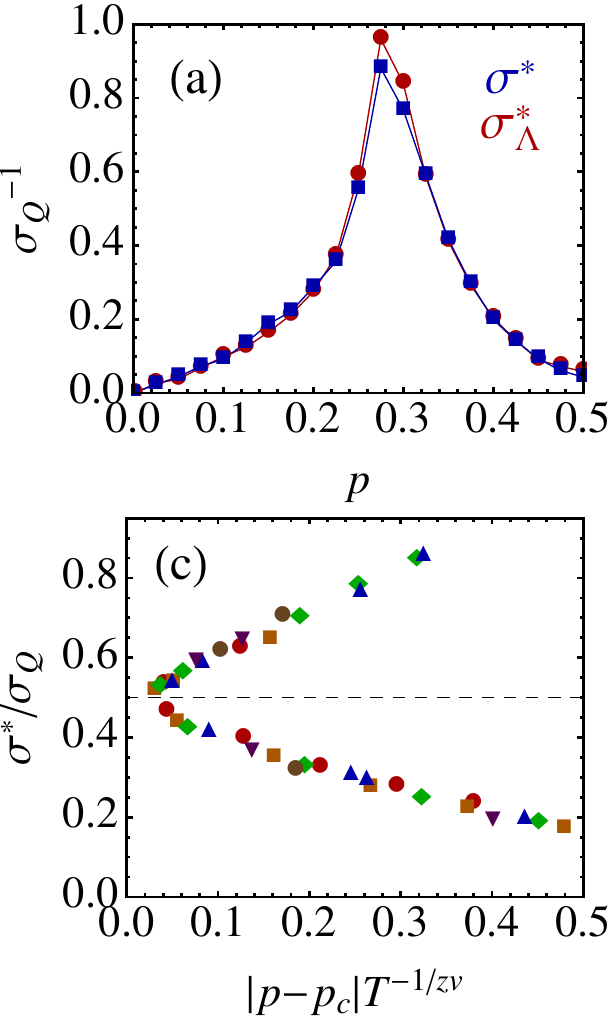}
\includegraphics[width= 0.225\textwidth]{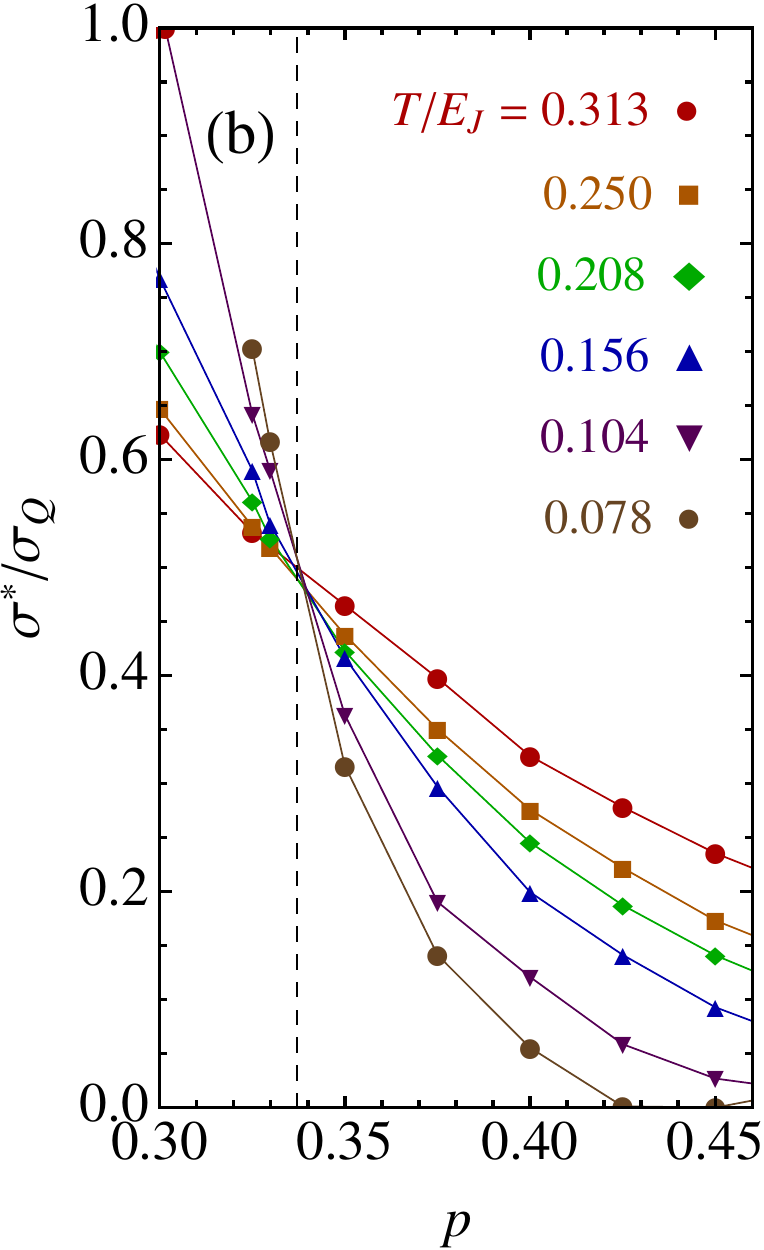}
}
\caption{	
(a) Comparison of two methods for obtaining the low-frequency conductivity near the SIT at $T/E_J = 0.156$, with
$\sigma^\ast$ from the integrated spectral weight in Eq.~(\ref{sigma-star}), and 
$\sigma^\ast_\Lambda$ from the current correlator $\Lambda_{xx}$ at imaginary time $\tau = \beta/2$ (see text).
(b) Plot of $\sigma^\ast(T; p)$ as a function of the disorder $p$ at various temperatures. The various curves cross at the
critical disorder strength $p_c$ at which $\sigma^\ast$ is $T$-independent with the critical value
$\sigma^\ast \approx 0.5\sigma_Q$. 
(c) Scaling collapse of the $\sigma^\ast(T; p)$ data with $p_c = 0.337$ and $z\nu = 0.96$, consistent with Fig.~\ref{QCP}.
\label{fig:sigmadc}
}
\end{figure}

As $E_c/E_J$ is tuned to reach the SIT in the clean system, $\rho_s$ decreases and vanishes at the transition; 
see Fig.~\ref{fig:energyscales}.
We also find that the Higgs scale goes soft upon approaching the quantum critical point, as expected.
%
%

The disordered SC results differ in several ways from those of the clean system. First,
the superfluid stiffness $\rho_s$ is reduced by disorder, 
vanishing at the SIT upon tuning the transition by disorder $p$.
An important difference is the absence of a discernible
Higgs threshold in $\Re \sigma(\omega)$ for the disordered SC; see Fig.~\ref{fig:dynamics}(b). 
Qualitatively we can understand this by the fact that once disorder breaks 
momentum conservation even single-phonon absorption is permitted
and one no longer needs a multi-phonon process for absorption. The effect of long-range Coulomb interactions, which change
the phonon dispersion ($\sim q$) to that of a 2D plasmon ($\sim \sqrt{q}$), is an important open problem. 
 
While the delta function in  $\Re \sigma(\omega)$ cannot be directly detected in dynamical
experiments, its Kramers-Kronig transform in the reactive response $\Im \sigma(\omega)=\rho_s/\omega$ can
indeed be measured. In the SC, the finite low-frequency absorption in $\Re \sigma(\omega)$ 
(due to the single-phonon processes discussed above)
causes 
$\omega \Im \sigma(\omega)$ 
to deviate from a constant,
as is evident in Fig.~\ref{fig:MEMresults}.
Our results are qualitatively similar to what has been seen in recent experiments, 
which, however, have focused on finite-temperature transitions in
weakly disordered samples~\cite{mondal2013}. 

\noindent
{\bf{Insulator:}}
The clean insulator shows a hard gap in $\Re \sigma(\omega)$ with an absorption threshold that we denote
by $\omega_\sigma$; see Fig.~\ref{fig:dynamics}(e). To gain insight into this gap, we look at the boson spectral function
$\Im P(\omega) / \omega$ in Fig.~\ref{fig:dynamics}(g), which too shows a hard gap $\omega_\text{B}$,
the analog of what was dubbed $\omega_{\rm pair}$ in Ref.~\onlinecite{bouadim2011}. The simplest process contributing 
to the conductivity is described diagrammatically as the convolution of two boson Greens functions leading
to $\omega_\sigma = 2\omega_\text{B}$ as seen in Fig.~\ref{fig:energyscales}.
We also see that both of these energy scales go soft as the SIT is approached from the insulating side.
In addition, there is a well-defined peak in $\Im P(\omega) / \omega$ at a characteristic scale $\widetilde{\omega}_\text{B}$
(Fig.~\ref{fig:dynamics}(g)), which also goes soft at the SIT (Fig.~\ref{fig:energyscales}). 

In contrast to the hard gap of the clean system, the dirty insulator exhibits absorption down to arbitrarily low frequencies
(see Fig.~\ref{fig:dynamics}(f)), which is, at least in part, due to rare regions. This then raises the question: what is
the characteristic energy scale that goes soft as one approaches the SIT from the insulating side? We find that this scale is
the location of the low-energy peak at $\widetilde{\omega}_\text{B}$ in the boson spectral function $\Im P(\omega) / \omega$, whose
evolution with disorder is most readily seen in the ``slingshot-like'' plot in Fig.~\ref{fig:MEMresults}(c). 
The corresponding changes in $\Re \sigma(\omega)$ are shown in Fig.~\ref{fig:MEMresults}(a). 
We also note that there is a marked change in $\Im \sigma(\omega)$ across the SIT.  We see from Fig.~\ref{fig:MEMresults}(b)
that it changes sign at low frequencies from an inductive ($\Im \sigma(\omega) > 0$) to a capacitive  ($\Im \sigma(\omega) < 0$) 
response going through the disorder-tuned SIT.

\noindent \textbf{Quantum criticality:}
We have already discussed the various scales that go soft on approaching the SIT from either side.
The results for the clean system, with SIT tuned by $E_c/E_J$, are summarized in Fig.~\ref{fig:energyscales}.
We now analyze the results for the disordered system.
The finite temperature QMC data, taken at face value, suggest a finite 
separation between the disorder values at which $\rho_s$ goes to
zero from the SC side and the characteristic boson scale $\widetilde{\omega}_\text{B}$
vanishes from the insulating side; see Fig.~\ref{QCP} (a,b). We emphasize that 
this intermediate region is {\it not a Bose metal} separating the SC and insulator, but rather the quantum critical region.
As shown in Fig.~\ref{QCP} (c), the SC transition temperature $T_c$, at which $\rho_s$ vanishes, and
the crossover scale $T^*$, at which  $\widetilde{\omega}_\text{B}$ vanishes, 
define this fan-shaped critical region. 
(We have used $z=1$ in scaling the system size as we go down in temperature in Fig.~\ref{QCP} (c); see Appendix A.) 
Both $T_c$ and $T^{*}$ extrapolate to zero at the same
critical disorder $p_c \approx 0.337$ (for the chosen value of $E_c/E_J$) with the scaling $|p - p_c|^{z\nu}$
where $z = 1$ and $\nu = 0.96 \pm 0.06$.

Finally, we turn to the important question of the universal conductivity at the SIT
\cite{cha1991,sorensen1992,cha1994,smakov2005,linSorensen2011}.
The d.c.~limit requires $\omega\!\to\!0$ first and then $T\!\to\!0$,
which is not possible when analytically continuing Matsubara data~\cite{damle1997}.
What we can meaningfully do is to
exploit quantum critical scaling and sum rules. 
The MEM results (i) satisfy the conductivity sum rule, which integrates over all
frequencies (see Appendix B), and (ii) are reliable for high frequencies $\omega > 2\pi T$.
Taking the difference of integrated spectral weights, we can reliably estimate 
$\sigma^\ast =  ({2 \pi T})^{-1}\int_{0^{+}}^{2 \pi T}\!d\omega  \Re\sigma(\omega,T;p)$.
We may now use the universal scaling form~\cite{damle1997} 
$\Re\sigma(\omega,T;p) = \sigma_Q\Phi\left(\omega/T;|p-p_c|T^{-1/z\nu}\right)$,
with $\sigma_Q=4e^2/h$,
to obtain
\be
\sigma^\ast(T;p) = {{\sigma_Q}\over {2 \pi}}\int_{0^+}^{2\pi} dx\, \Phi(x;|p-p_c|T^{-1/z\nu}).
\label{sigma-star}
\ee
Thus $\sigma^\ast$ is a $T$-independent universal constant at the quantum critical point $p=p_c$ and closely related to the low frequency conductivity measured in experiments.

Another estimate of the low-frequency conductivity comes directly from the current correlator
$\sigma^\ast_\Lambda = \beta^2 \Lambda_{xx}({\bf q}=0,\tau = \beta/2)/{\pi}$
at the largest available value of imaginary time~\cite{trivedi1996}. The
$\sigma^\ast$ estimates obtained by the two methods show
good agreement (Fig.~\ref{fig:sigmadc}(a)) and provide a non-trivial check on the analytic continuation.

In Fig.~\ref{fig:sigmadc} (b) we plot $\sigma^\ast(T; p)$ as a function of $p$ for various temperatures. 
In the superconductor ($p<p_c$) the conductivity increases with decreasing $T$, 
while the opposite trend is observed in the insulator ($p>p_c$). 
Precisely at the SIT $p=p_c$, we find a $T$-independent crossing point which also allows us
to estimate the critical $\sigma^\ast$. Another way to scale the data is to plot $\sigma^\ast(T; p)$
as a function of the scaling variable $|p-p_c|T^{-1/z\nu}$. We find data collapse for $p_c = 0.337$ and $z\nu = 0.96$
(consistent with Fig.~\ref{QCP}) with a critical value of $\sigma^\ast\approx 0.5 \sigma_Q$. 
For a detailed comparison of the critical exponents and $\sigma^*$ with previous results~\cite{iyer2012}, see Appendix C.

\noindent \textbf{Conclusions:} 
We have presented calculations of the complex dynamical conductivity $\sigma(\omega) $ and the boson spectral function $P(\omega)$
across the SIT driven by increasing the charging energy $E_c/E_J$ as well as by increasing disorder $p$.
By comparison of the clean and disordered problems, we see the effect of disorder on the Higgs scale $\omega_{\rm Higgs}$ in the superconductor and on the threshold 
$\omega_\sigma$ in the insulator, in generating low frequency weight in absorption in both superconducting and insulating phases, 
and in expanding the region over which critical fluctuations are observable.
In the literature, an insulating phase of bosons with disorder has been referred to as a compressible Bose glass phase away from particle-hole symmetry~\cite{fisher1989} or an incompressible Mott glass phase with particle-hole symmetry~\cite{iyer2012}. We work with a particle-hole symmetric system, and while we see evidence of a gap-like scale in the insulator, we also find a low-frequency tail in the absorption, presumably arising from rare regions.
In this respect our insulator seems more akin to a Bose glass.
It is important to emphasize that the effects we have calculated have required going beyond mean field theories, even those that included emergent granularity due to the microscopic disorder, by focussing on the role of fluctuations of the order parameter.
We have calculated the effect of these fluctuations, both amplitude and phase, on experimentally accessible observables using QMC methods coupled with maximum entropy methods, constrained by sum rules. Recently the AdS-CFT holographic mapping has been used to obtain the dynamical conductivity at the 
disorder-free bosonic quantum critical point~\cite{witczak2013,chen2013universalconductivity}.
Our focus here has been on the evolution of the dynamical quantities in both the phases, superconducting and insulating, and across the 
disorder-driven SIT, for which the holographic formalism has not yet been developed.  

Our calculations have laid the foundation for key signatures in dynamical response functions 
across quantum phase transitions.
Though we have focused on the disorder-driven s-wave SIT in thin films, the ideas are 
equally relevant for a diverse set of problems, including: 
(i) unconventional superconductors like the high Tc cuprates that have a quantum critical point tuned by doping,
(ii) SIT at oxide interfaces like LaAlO$_3$/SrTiO$_3$,
(iii) SIT in the next generation of weakly coupled layered materials like dichalcogenide monolayers, and
(iv) bosons in optical lattices with speckle disorder.

\section*{ACKNOWLEDGEMENTS}
We thank Assa Auerbach, Subir Sachdev, and William Witczak-Krempa for discussions. 
We gratefully acknowledge support from an NSF Graduate Research Fellowship (M.S.),
DOE DE-FG02-07ER46423 (N.T.), NSF DMR-1006532 (M.R.), and
computational support from the Ohio Supercomputing Center.
MR and NT acknowledge the hospitality of the Aspen Center for Physics,
supported in part by NSF PHYS-1066293.


\bigskip

\section*{Appendix A: Monte Carlo simulations}


We analyze the (2+1)D quantum XY model given by Eq.~\ref{ham}, which is a generalization of the full quantum rotor Hamiltonian
\begin{eqnarray} \nonumber
\hat{H}_{J} &=& \frac{E_c}{2} \sum_i{\hat{n}_i}^2 
 - \sum_{\lab i j \rab} J_{ij} \cos{(\hat{\theta}_i- \hat{\theta}_j)} 
 - \sum_{i} \left( \mu - V_i \right) n_i \\
 &=& -\frac{E_c}{2} \sum_i{\frac{d^2}{d\theta_i^2}} 
 - \sum_{\lab i j \rab} J_{ij} \cos{(\hat{\theta}_i- \hat{\theta}_j)} \\ \nonumber
 && \  + \ i \sum_{i} \left( \mu - V_i \right) \frac{d}{d\theta_i}
\end{eqnarray}
since $n_i = - i d/d\theta_i$.
The partition function can be expressed as the coherent-state path integral $Z = \int D[\theta] e^{-S}$ with action~\cite{cha1991}
\begin{eqnarray} \nonumber
S &=& \int_0^\beta d\tau \Big\{ \frac{1}{E_c} \sum_i ({\partial_\tau} \theta_i)^2  - i \frac{\mu - V_i}{E_c/2} \partial_\tau \theta_i \\
&& \ \ \ \ \ \ \ \ \ \ \ \ \ \ \ \ \ \ \ \ - \sum_{\lab i j \rab} J_{ij} (1 - \cos[(\hat{\theta}_i- \hat{\theta}_j)]) \Big\}.
\end{eqnarray}
For a slowly varying phase, this becomes
\begin{equation}
S = \int_0^\beta d\tau \Big\{ \frac{1}{E_c} \sum_i ({\partial_\tau} \theta_i)^2  - i \frac{\mu - V_i}{E_c/2} \partial_\tau \theta_i - \frac{J_{ij}}{2} (\partial_r \theta_i)^2 \Big\}.
\end{equation}

For the pure system ($V_i = V$), if $(\mu - V)/(E_c/2)$ is an integer, then the middle term does not contribute to the free energy because $\int_0^\beta \partial_\tau \theta_i = 2 \pi \times \rm{integer}$.
In this special case of particle-hole symmetry, the dynamical exponent is $z = 1$.
Away from this particle-hole symmetry point, the first derivative term remains in the action, and for the pure system $z = 2$.
Upon including disorder $V_i$ in the diagonal potential, recent Monte Carlo simulations~\cite{meier2012} have obtained  $z = 1.83 \pm 0.05$.

The model we have studied has bond disorder $J_{ij}$  that respects particle-hole symmetry.
In this case the dynamical exponent is expected to remain z  = 1 as argued in Refs.~\onlinecite{fisher1991} and ~\onlinecite{weichman2008}.
We also note that a recent Monte Carlo study of the (1+1)D JJA also concluded that $z  = 1$ in the presence of bond disorder~\cite{hrahsheh2012}.
 In order to definitively establish the value of $z$, two-parameter finite-size scaling with varying aspect ratios of $L_\tau$ and $L$ is necessary. Within the scope of the analysis presented here the good scaling collapse of our data for the bond-disordered model, shown in Figs.~\ref{QCP} and~\ref{fig:sigmadc}, is indeed consistent with $z  = 1$.

Our Monte Carlo simulations are performed by mapping it Eq.~\ref{ham} onto
an anisotropic 3D classical XY model with Hamiltonian~\cite{cha1994}
\begin{equation}
\begin{split}
H_{\textrm{XY}} &= - K_{\tau} \sum_{\vec{r}, \ j} \cos{[\theta_{\vec{r}}(\tau_j) - \theta_{\vec{r}}(\tau_{j+1})]}  \\
                          & \ \ \  - K_0 \sum_{\lab \vec{r},\vec{r'} \rab, \ j} \cos{[\theta_{\vec{r}}(\tau_j)- \theta_{\vec{r'}}(\tau_j)]}
\label{XY}
\end{split}
\end{equation}
by performing a Trotter decomposition of imaginary time into $L_{\tau}$ slices of width $\Delta \tau$ such that the inverse temperature $\beta = L_\tau \Delta \tau$;
$\vec{r}$ and $\vec{r'}$ are points in the 2D plane and $\tau_j$ denotes the $j^{th}$ imaginary time slice;
and the dimensionless coupling constants are $K_{\tau} = 1/ \Delta \tau E_c$ and $K_0 = \Delta \tau E_J$.

We perform Monte Carlo simulations using the efficient Wolff cluster update method~\cite{wolff1989}.
In all of our simulations, we have set $K_0 = 0.1$, which we have checked to be sufficiently small to remove the error from the Trotter decomposition.
For the clean system, we performed simulations on lattices of size $ 256 \times 256$ with $L_\tau = 64$.
For the disorder tuned transition, we worked at fixed $E_c/E_J = 3.0$. 
Simulations at different temperatures have been performed by changing the number of imaginary time slices 
$L_\tau$ from 32 to 128; for each $L_\tau$, we fix $L=L_\tau$ since the dynamical exponent is $z=1$. 
All disorder results have been averaged over 100 disorder realizations.

\section*{APPENDIX B: Dynamical observables and analytic continuation}
We calculate the imaginary time, or equivalently the Matsubara frequency ($\omega_n = 2 n \pi /\beta$), 
current-current correlation function 
\be
\Lambda_{xx}(\mathbf{q}; i \omega_n) = \sum_{\mathbf{r}}\int_0^\beta\!{d\tau \langle j_x(\mathbf{r}, \tau) j_x(0,0) \rangle e^{i \mathbf{q} \cdot \mathbf{r}} 
e^{-i\omega_n \tau}}
\ee
where the paramagnetic current in our model is given by $j_x(\vec{r},\tau) \equiv K_0 \sin{[\theta(\vec{r} + \hat{x},\tau) - \theta(\vec{r} ,\tau)]}$.
The conductivity is related to the analytic continuation of $\Lambda_{xx}$ at ${\bf q}=0$ 
\be \label{sigma}
\sigma(\omega) = \left[\kx - \Lambda_{xx}(\omega + i0^+)\right]/i (\omega + i0^+)
\ee
where $\langle - k_x \rangle$ is the average kinetic energy along bonds in the $x$-direction.
$\Re \sigma(\omega)$ is then given by
\be \label{sigmaEq}
\Re  \sigma(\omega) = \rho_s \delta(\omega) + \Im  \Lambda_{xx}(\omega)/{\omega}.
\ee
The superfluid stiffness $\rho_s$ is obtained from the difference between the transverse and longitudinal 
limits of the current-current correlation function 
$\rho_s / \pi = \Lambda_{xx}(q_x \rightarrow 0, q_y = 0, i\omega_n=0) - \Lambda_{xx}(q_x = 0, q_y \rightarrow 0, i\omega_n=0)$
and the sum rule  $\langle - k_x \rangle = \Lambda_{xx}(q_x \rightarrow 0, q_y = 0, i\omega_n=0)$.
Finally, $\Re\sigma(\omega)$ obeys the
optical conductivity sum rule $2\int_{0^+}^\infty d\omega \Re\sigma(\omega) = \pi\langle - k_x \rangle - \rho_s$,
which serves as a non-trivial check on our analytic continuation results.

%
%

{\bf (1) Analytic continuation of $\Lambda(\tau)$}:
The imaginary time correlation function $\Lambda_{xx}(\tau)$ calculated in our Monte Carlo simulations is related to its real-frequency counterpart (and subsequently to $\sigma(\omega)$) through 
\begin{equation} \label{LT}
\Lambda_{xx}(\tau) = \int_{- \infty}^{\infty} \frac{d\omega}{\pi} \ \frac{e^{-\omega \tau}}{1 - e^{-\beta \omega}} \Im \Lambda_{xx}(\omega).
\end{equation}
To extract the real frequency data, we have employed the maximum entropy method (MEM)~\cite{Gubernatis1991} to invert this Laplace transform.
We have performed extensive tests on our Maximum Entropy routine; further details can be found in the supplemental material of Ref. \onlinecite{bouadim2011}.

\begin{figure}[!t]
\includegraphics[width= 0.45 \textwidth]{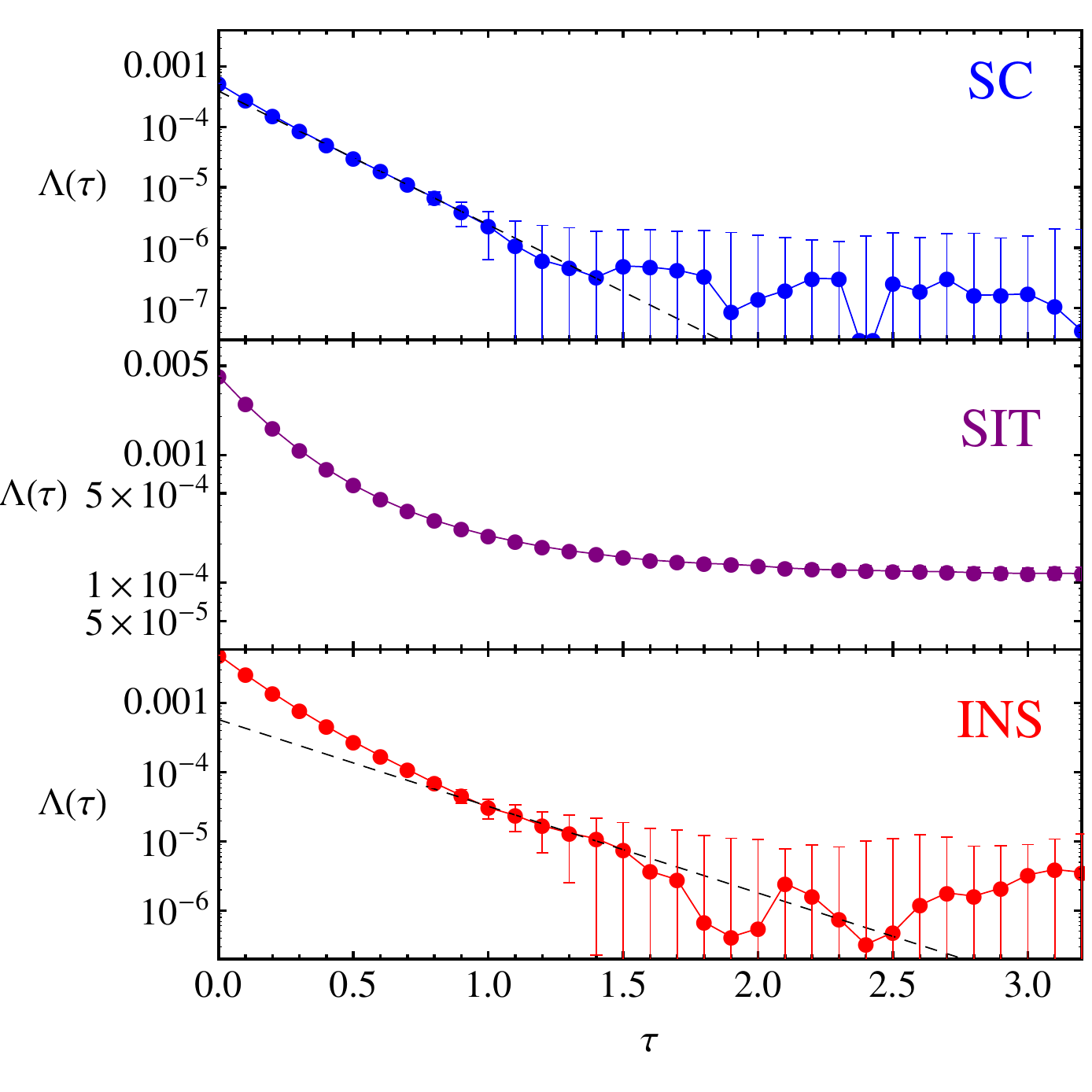}
\caption{	
	Imaginary time correlation functions $\Lambda(\tau)$ for the clean transition in the superconducting phase ($E_c/E_J = 3.22$), insulating phase ($E_c/E_J = 4.76$), and near the critical point ($E_c/E_J = 4.17$).
	Dashed lines indicate the gap scales that can be reliably extracted in the SC and insulating phases.
	}
\label{fig:cleanlambdatau}
\end{figure}

In addition to these tests on the MEM routine itself, whenever possible, we have checked those characteristics of the spectra obtained via the MEM against features which can be directly calculated for the Monte Carlo correlation functions. 
Gapped functions, either in $\sigma(\omega)$ and $P(\omega)$, have recognizable exponential decays in the imaginary time correlation functions corresponding to the gap scale, whereas spectra without a gap correspond to correlation functions with no discernible gap scale in the $\tau$ data, see Fig.~\ref{fig:cleanlambdatau}.
The extracted gap scales from $\Lambda(\tau)$ or $P(\tau)$ track consistently with those scales coming from the real frequency functions obtained after performing the analytic continuation.
This is true both in the clean system and in the disordered system, where the presence of even small disorder makes the reading off a Higgs scale in the superconductor unreliable, consistent with the analytically continued results.

\begin{figure}[!t]
\centerline{\includegraphics[width= 0.25 \textwidth]{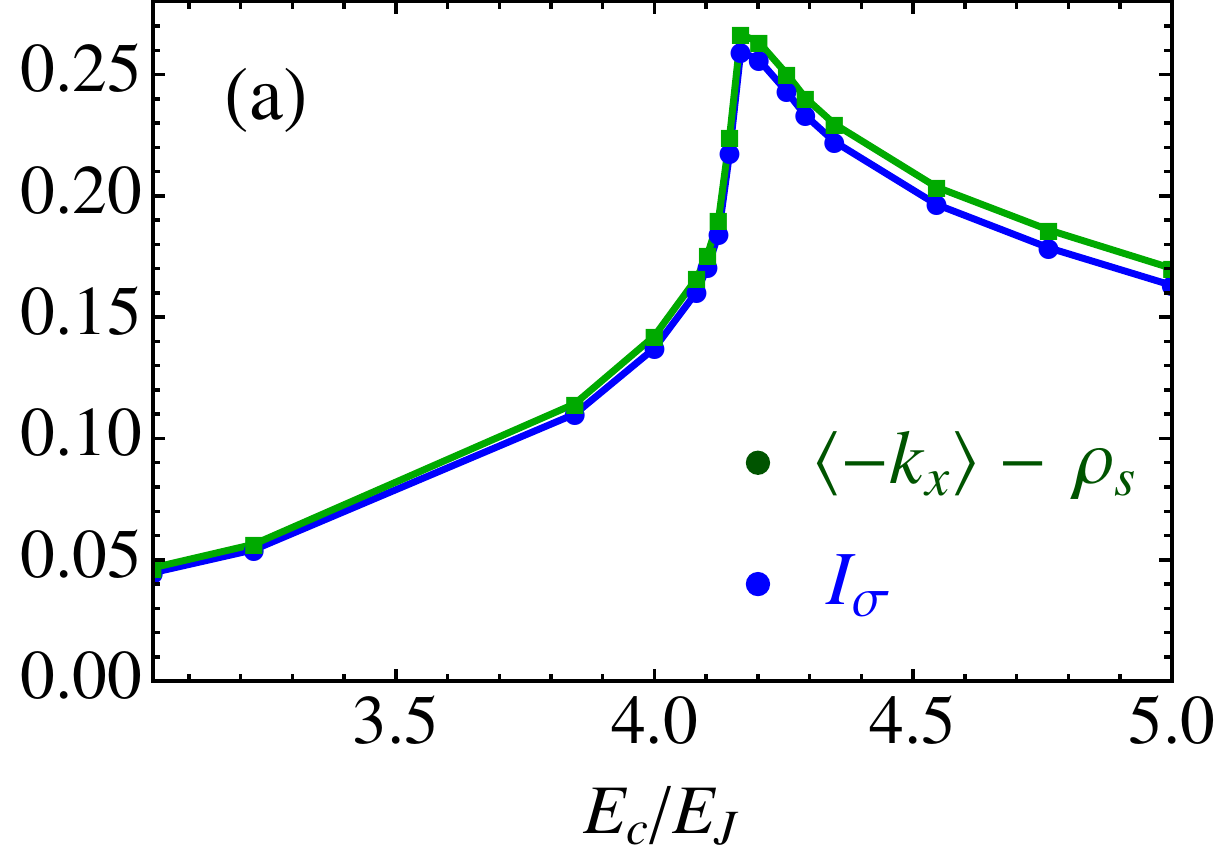}
\includegraphics[width= 0.25 \textwidth]{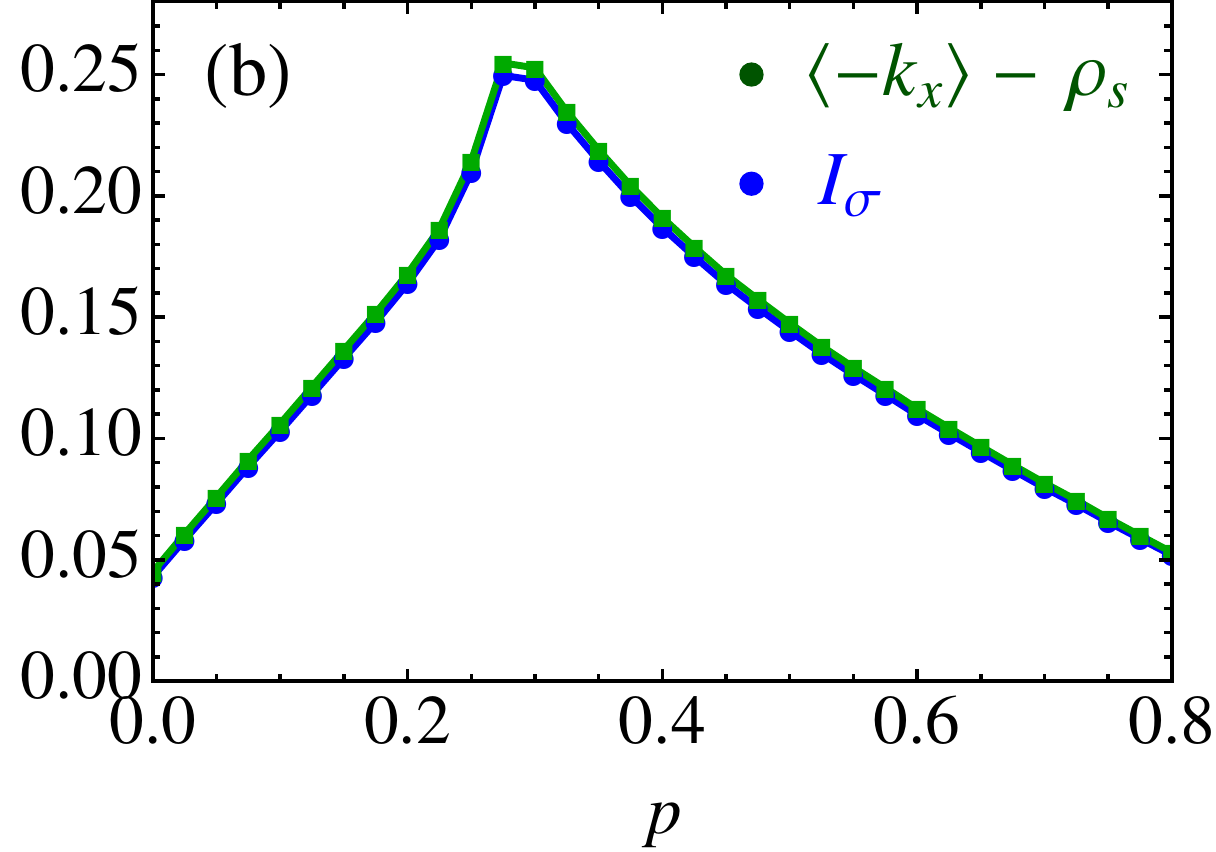}}
\centerline{\includegraphics[width= 0.25 \textwidth]{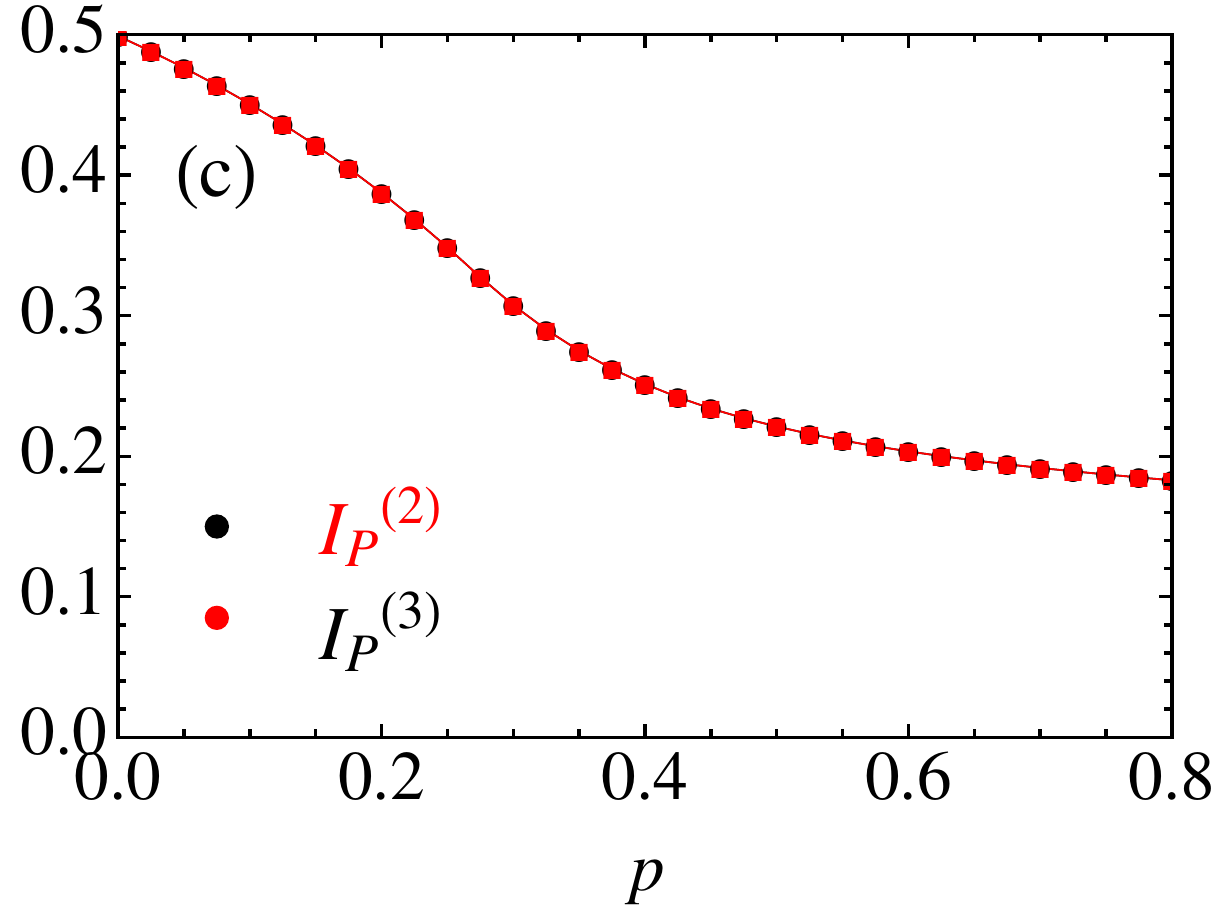}
\includegraphics[width= 0.25 \textwidth]{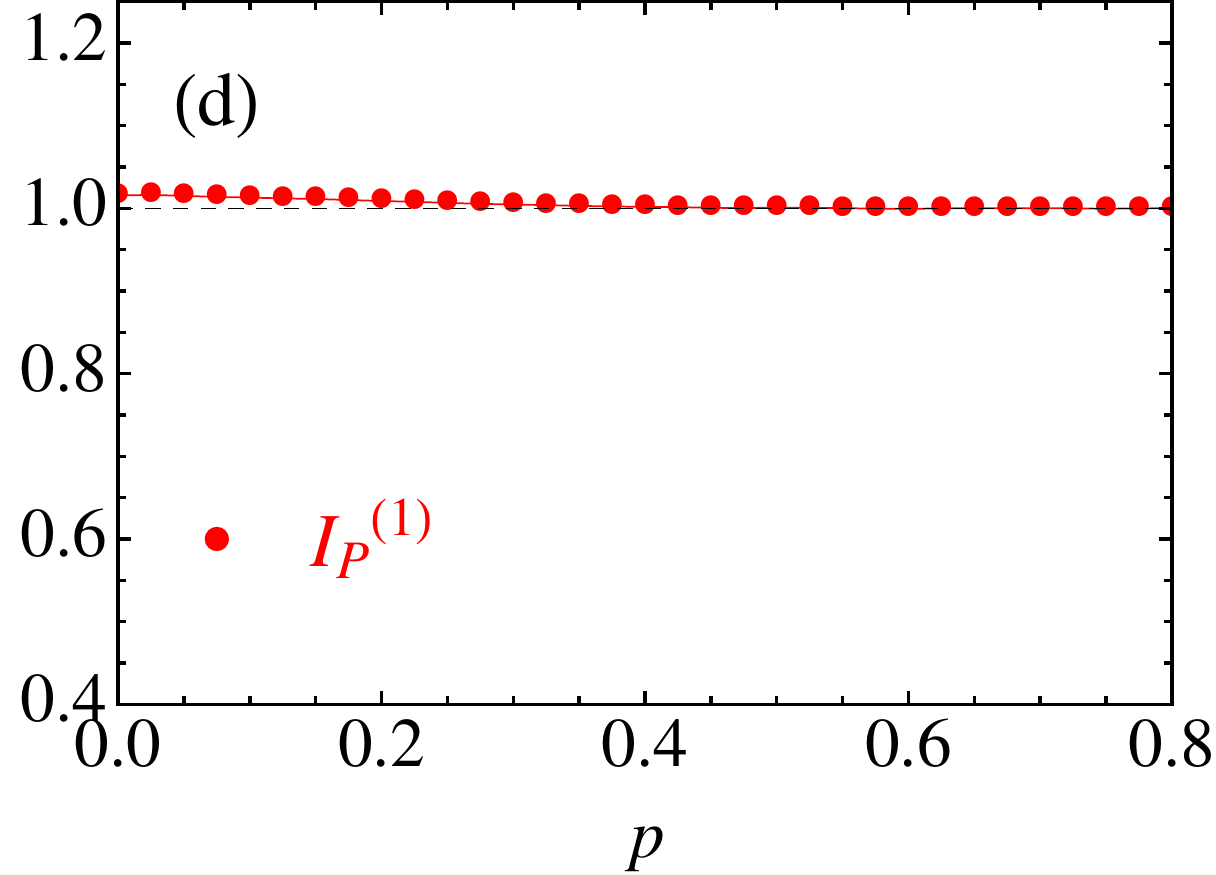}}
\caption{	
	Sum rules for quantities calculated using Maximum Entropy analytic continuation.
	(a) and (b) show the conductivity sum rule Eq.~\ref{sigmasumrule} for the clean ($p = 0$) and disorder tuned ($E_c/E_J$) transition.
	(c) and (d) show the sum rules given by Eqs.~\ref{sumonP} and~\ref{sumonPoverw} for the boson spectral function $P(\omega)$for the disordered tuned transition.
	In all cases, the results shown are for $T/E_J = 0.156$, but hold at all temperatures considered in this work.
	}
\label{fig:sumrules}
\end{figure}

We have also carefully checked that the sum rule on $\sigma(\omega)$ is verified.
For lattice systems, the optical conductivity sum rule is
\be \label{sigmasumrule}
I_\sigma = \int_{-\infty}^{\infty} d\omega \ \Re \sigma(\omega) = \pi \lab-k_x\rab.
\ee
This includes the spectral weight contained in the delta-function response proportional to the superfluid stiffness $\rho_s$.
The regular part of the spectrum (which we obtain from analytic continuation) satisfies
\be
2 \int_{0^+}^{\infty} d\omega \ \Re \sigma(\omega) = \pi \lab-k_x\rab - \rho_s.
\ee
We emphasize that this sum rule is not built into the MEM routine, and provides an independent verification of the procedure.
The sum rule is shown in Fig.~\ref{fig:sumrules} for both the clean and the disorder driven transitions.

{\bf (2) Analytic continuation of $P(\tau)$}:
The boson spectral function is related to the boson Greens function via
\begin{equation} \label{Panalcont}
P(\tau) = \int_{- \infty}^{\infty} \frac{d\omega}{\pi} \ \frac{e^{-\omega \tau}}{1 - e^{-\beta \omega}} \Im P(\omega)
\end{equation}
which we invert using the MEM in exactly the same way as for $\sigma(\omega)$.

The boson spectral functions obeys the following two sum rules
\be \label{sumonP}
I_P^{(1)}=\int_{-\infty}^{\infty} \frac{d\omega}{\pi} \frac{1}{1-e^{-\beta \omega}} \Im P(\omega) = P(\tau=0) = 1,
\ee
which follows trivially from Eq.~\ref{Panalcont}, and 
\be \label{sumonPoverw}
I_P^{(2)}=\int_{-\infty}^{\infty} \frac{d\omega}{\pi} \frac{\Im P(\omega)}{\omega} = \int_0^\beta d\tau P(\tau) = I_P^{(3)}.
\ee
This second sum rule can be seen by integrating both sides Eq.~\ref{Panalcont} over $\tau$ from $0$ to $\beta$. Note that, in the limit of $T \rightarrow 0$, Eq.~\ref{sumonP} reduces to a sum rule on $\Im P(\omega)$ itself $\int_0^{\infty} d\omega \Im P(\omega) = 1$.
Results for the sum rules are shown in Fig.~\ref{fig:sumrules}.

\section*{APPENDIX C: Universal conductivity and critical exponents}

There have been many attempts to calculate the value of the so-called universal conductivity at the superconductor-insulator quantum phase transition.
We will only focus on those models expected to be in the same universality class as our model ($z=1$);
a more complete history can be found in Ref.~\onlinecite{lin2011} and the references therein.

Since the conductivity is a universal function of $\omega/T$ at the critical point~\cite{damle1997}, there are different and possibly distinct limiting values of $\sigma(\omega/T)$
\bea
&\sigma(0) &= \sigma(\omega \rightarrow 0, T = 0) \\
&\sigma(\infty) &= \sigma(\omega = 0, T \rightarrow 0)
\eea
In our paper, we have proposed another universal quantity
\be
\sigma^\ast = \frac{\sigma_Q}{2 \pi T} \int_{0^+}^{2\pi T} d\omega\ \sigma(\omega)
\label{sigma-star}
\ee
that can be reliably extracted from the numerics as explained in the text.

We will express all $\sigma$ values in units of $\sigma_Q = 4e^2/h$.
For disorder-free models in the (2+1)D XY universality class, our value of $\sigma^* \approx 0.4$ is consistent with the recent result of Ref.~\onlinecite{smakov2005}, where they found $\sigma(0) = 0.45 \pm 0.05$ using Pad\'{e} approximates to analytically continue MC data at the critical point, modified from the previous estimate~\cite{cha1991} of $\sigma(0) = 0.285 \pm 0.02$ obtained by extrapolation of the current-current correlation function for $\omega_n \rightarrow 0$.
More recently, groups~\cite{witczak2013,chen2013universalconductivity} have used holographic continuation to perform analytic continuation at the critical point. They find $\sigma(\infty) = 0.32$ and $\sigma(\infty) = 0.359(4)$ respectively.

For the disorder-tuned transition, we have obtained
\bea
&\sigma^* &\approx 0.50 \\
&\nu &= 0.96 \pm 0.06
\eea
where $z=1$ by definition for the model.
There have been only a few results on disordered transitions that can meaningfully be compared to our work.
A Monte Carlo study of the (2+1)D XY model with onsite charging energy disorder~\cite{cha1994} found $z = 1.07 \pm 0.03$, $\nu \approx 1$, and $\sigma(0) = 0.27 \pm 0.04$ obtained by extrapolation of $\Lambda_{xx}$ for $\omega_n \rightarrow 0$.
We expect that using analytic continuation could modify this estimate.
Studies of the disordered quantum rotor model using strong disorder renormalization group theory~\cite{iyer2012} have found $\nu = 1.09 \pm 0.04$, although they have not looked at the universal conductivity.


\end{document}